\documentclass[a4paper, 12pt]{article}

\usepackage{amsmath}
\usepackage{amssymb}
\usepackage{feynmp-auto}
\usepackage{dcolumn}
\usepackage{blindtext}
\usepackage{multicol}
\usepackage{bbm}
\usepackage{mathrsfs}
\usepackage{units}
\usepackage{graphicx}
\usepackage{todonotes}
\usepackage{slashed}

\usepackage{cancel}
\usepackage[scr=boondox,  
            ]   
           {mathalpha}

\graphicspath{{./plots}}

\usepackage{jheppub} 
\title{Measuring lepton number violation in heavy neutral lepton decays at the future muon collider}
\author[1]{Oleksii~Mikulenko,}
\emailAdd{mikulenko@lorentz.leidenuniv.nl}
\author[1,2]{Mariia~Marinichenko}
\emailAdd{marinichenko.m@umail.leidenuniv.nl}
\affiliation[1]{Instituut-Lorentz, Leiden University, Niels Bohrweg 2, 2333 CA Leiden, The Netherlands}
\affiliation[2]{Department of Physics, Taras Shevchenko National University of Kyiv, Volodymyrska Street 64/13, 01601 Kyiv, Ukraine}

\date{}
\begin{document}

\abstract{
The future muon collider has the potential to discover feebly interacting particles in a wide range of masses above the electroweak scale. It is particularly suitable to search for heavy neutral leptons (HNLs), as their production cross section $\sigma \sim m_W^{-2}$ is not suppressed by the new physics scale. We demonstrate that with the capacity to observe up to $10^5$ events in the previously unexplored TeV mass range, the muon collider provides the means to measure the fraction of lepton number violating (LNV) processes with precision at the level of a percent. This capability enables elucidating the nature of HNLs, allowing us to differentiate between Majorana, Dirac, and non-minimal scenarios featuring multiple degenerate HNLs. We link the observed fraction of LNV processes to the parameters of the model with three degenerate HNLs, which could be responsible for generating baryon asymmetry in the Universe. Additionally, we present a simple estimate for the number of signal events, as well as analyze the feasibility of vector boson fusion processes in searches for HNLs.
}

\maketitle

\section{Introduction}

A few puzzles remained to be addressed by the Standard Model (SM) of particle physics, such as the origin of neutrino masses and matter-antimatter asymmetry in the Universe. A notable solution is to complete the SM with the right-handed counterparts of the left-handed neutrinos - heavy neutral leptons (HNL). With such a minor adjustment, it is possible to explain the current neutrino oscillation data through the seesaw mechanism~\cite{ParticleDataGroup:2020ssz, Minkowski:1977sc, Schechter:1980gr, Schechter:1981cv, Gell-Mann:1979vob, Yanagida:1979as, Mohapatra:1979ia}, while simultaneously providing the HNLs with a neutrino-like weak interaction, suppressed by a tiny mixing angle $U_\alpha^2\ll1$, $\alpha = e$, $\mu$, $\tau$. At the same time, the new particles may produce the observed amount of baryon asymmetry in the Universe~\cite{Boyarsky:2009ix, Asaka:2005an, Canetti:2012vf, Shuve:2014zua, Abada:2015rta, Hernandez:2015wna, Drewes:2016gmt, Hernandez:2016kel, Hambye:2017elz, Abada:2017ieq, Antusch:2017pkq, Klaric:2020phc, Klaric:2021cpi, Drewes:2016jae, Canetti:2012kh} or be a portal into a more complex dark sector with successful baryogenesis~\cite{Hall:2019ank, Hall:2019rld, Hall:2021zsk}. Both these goals may be achieved with GeV-TeV scale HNLs that can be probed by the current and proposed future experiments~\cite{Alekhin:2015byh, SHiP:2015vad, Aberle:2839677, Baldini:2021hfw, Alviggi:2839484, CortinaGil:2839661, DUNE:2020ypp, DUNE:2020fgq, Batell:2020vqn, MATHUSLA:2019qpy, Cerci:2021nlb, FASER:2018bac, SHiP:2020sos, SNDLHC:2022ihg, Bauer:2019vqk, Aielli:2019ivi, Dercks:2018wum, Boyarsky:2022epg, Chrzaszcz:2020emg, Antusch:2016ejd}.

The project of a muon collider~\cite{Delahaye:2019omf, AlAli:2021let} is a possible way forward to explore physics at the TeV scale, an alternative to future $pp$ and $ee$ colliders: Future Circular Collider (FCC)~\cite{Agapov:2022bhm, Benedikt:2022kan}, Circular Electron-Positron collider (CEPC)~\cite{CEPCPhysicsStudyGroup:2022uwl}, Compact Linear Collider (CLIC)~\cite{Linssen:2012hp}, and International Linear Collider (ILC)~\cite{Behnke:2013xla}. Muons can be accelerated to high energies up to $\unit[1 - 10]{TeV}$, equivalent to a $\unit[10-100]{TeV}$ proton collider~\cite{Delahaye:2019omf, Costantini:2020stv}. At the same time, muons are fundamental particles and provide a clean environment, allowing for a simple description in high-energy scatterings. Finally, new physics in the lepton sector may be probed more efficiently in processes with muons, as compared to hadron colliders.

We consider the two standard setups with the energies $\sqrt{s} = 3$, $\unit[10]{TeV}$ and integrated luminosities $\mathcal L = 1$, $\unit[10]{ab^{-1}}$, respectively. Throughout the paper, we assume that the muon beams are fully polarized, with muons having negative helicity and antimuons having positive helicity correspondingly. The detector apparatus coves the angles above $\theta_\text{det} = 10^\circ$ relative to the beam axis.
  
In these setups, the muon collider may be a powerful instrument in searching for feebly interacting HNLs in the region above the electroweak scale. The reason behind it relies on the following considerations. The naive scaling of the cross-section for processes involving a new physics particle of mass $M$, either produced or appearing as an off-shell mediator, is
\begin{equation}
\label{eq:sigma_s}
    \sigma_M \sim \frac{g^4}{16\pi M^2} = \unit[10^2]{fb} \times g^4 \left(\frac{\unit[10]{TeV}}{M}\right)^2,
\end{equation}
where $g$ is the coupling constant of the considered interaction. The luminosity for different runs is assumed to scale as $\mathcal L \propto s$ to account for the suppression of the cross-section at higher energies. The corresponding number of events is
\begin{equation}
    \mathcal L \sigma \sim 10^6 g^4 \sim \unit[10^4]{events}
\end{equation}
for $g^2\sim 0.1$ (electroweak coupling). However, this is not the case for TeV-scale HNLs, which are produced mainly in the $t$ channel process, mediated by the $W$-boson~\cite{Antusch:2016ejd}. The propagator that may carry a light-like transferred momentum $q^2 \to 0$ introduces a divergence in the differential cross-section, which is regularized by the mediator mass. Therefore, the cross-section avoids the suppression by the new physics scale and is enhanced to
\begin{equation}
    \sigma \sim U_\mu^2\sigma_\text{weak}, \qquad \sigma_{\text{weak}} \sim \frac{g^4}{16 \pi m_W^2} \sim \unit[200]{pb}
\end{equation}
with the expected number of produced HNLs being
\begin{equation}
    \label{eq:NHNL_estimate}
    N_\text{ev} \sim U_\mu^2  \mathcal  L \sigma_{\text{weak}} \sim \frac{U_\mu^2}{10^{-8}} \times \frac{\mathcal L}{\unit[10]{ab^{-1}}}\,\text{events}.
\end{equation}

\begin{figure}[h!]
    \centering
    \includegraphics[width = 1.0\textwidth]{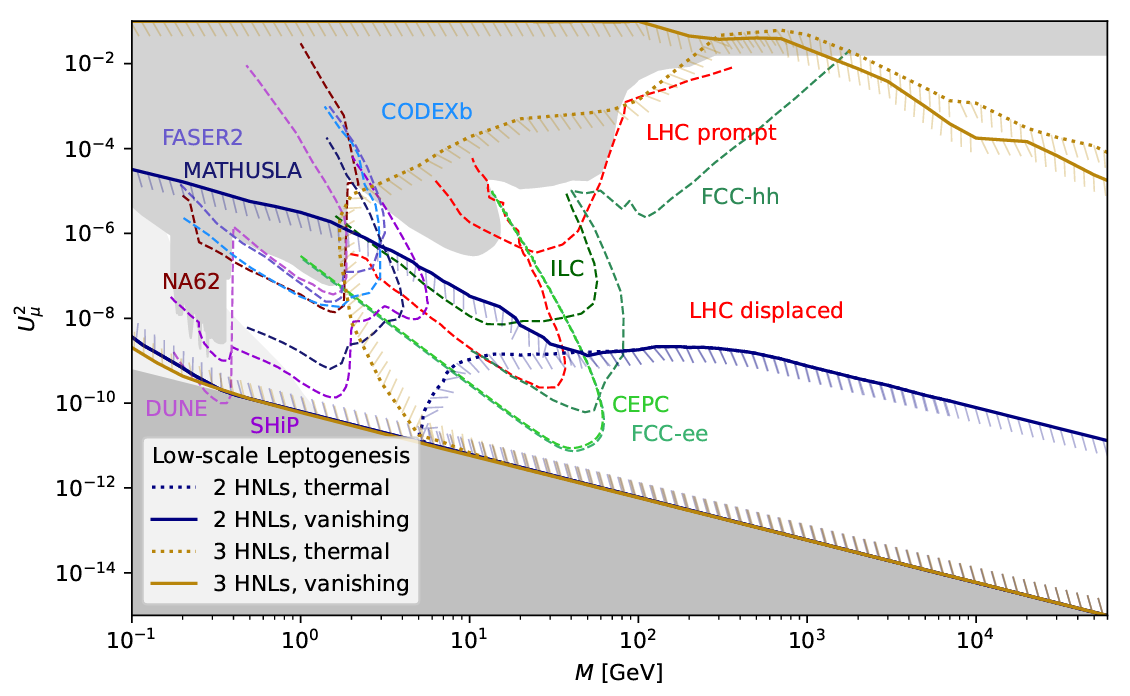}
    \caption{Constraints on heavy neutral leptons from~\cite{Abdullahi:2022jlv}, see reference therein. The shaded areas are excluded regions by direct detection and electroweak precision measurements (above) or seesaw limit (below). The parameter space with HNL mass above $\unit[100]{GeV}$ is consistent with 3 HNL leptogenesis~\cite{Drewes:2021nqr} and is a target for searches at multi-TeV colliders.  }
    \label{fig:leptogenesis}
\end{figure}

The studies of the sensitivity to HNLs of an experiment at the muon collider have recently been performed in~\cite{Kwok:2023dck, Li:2023tbx, Mekala:2023diu}, with the obtained sensitivity down to $U^2_\mu \sim 10^{-6}$. The potential to explore HNLs above the electroweak scale with many events provides a unique opportunity to study the properties of the new particles. The possibility of inferring the underlying physics model of HNLs, if detected, and linking it to the origin of neutrino masses and baryon asymmetry has attracted significant attention~\cite{Antusch:2017pkq, Tastet:2019nqj, Tastet:2021vwp, Tastet:2021ygq, Drewes:2018gkc, Bondarenko:2021cpc}. In the minimal leptogenesis scenario with two quasi-degenerate HNLs, the range of the mixing angles is beyond the expected sensitivity reach. However, the existence of the third HNL may significantly alter the dynamics of the production of matter-antimatter asymmetry, yielding a much larger parameter space~\cite{Abada:2018oly, Drewes:2021nqr}, which may be probed by an experiment hosted at the muon collider, see Fig.~\ref{fig:leptogenesis}.

In this paper, we discuss the possible implications of the muon collider to reveal the underlying properties of the HNLs. Namely, we analyze the possibility of probing the leptogenesis scenario with 3 HNLs. In Sec.~\ref{Sec:2}, we complement the previous simulation-based result with simple analytical estimates for the signal yield for HNLs. Additionally, we discuss caveats and limitations associated with probing HNL couplings to $e$, $\tau$ lepton flavors by vector boson fusion (VBF). In Sec.~\ref{Sec:3}, we obtain the sensitivity to observations of lepton number violation and the implications of such measurement to a model of 3 HNLs. We discuss our findings in Sec.~\ref{Sec:conclusion}.

\section{Analytic estimate of the sensitivity reach}
\label{Sec:2}

In the framework of the type-I seesaw mechanism, the Lagrangian with $n$ right-handed neutrinos has the following form \cite{Abdullahi:2022jlv}:
\begin{equation}
\label{eq:HNL_Lagrangian}
    \mathcal{L}=\mathcal{L}_{SM}+i\overline{N}_{I}\gamma^{\mu}\partial_{\mu}N_{I}-F_{\alpha I}\overline{L}_{\alpha}N_{I}\Tilde{H}-\frac{1}{2}M_{IJ}\overline{N}^{C}_{I}N_{J}+h.c.,
\end{equation}
where $N_{I}$, $I= \overline{1,n}$ are HNL states with Majorana mass matrix $M_{IJ}$; $F_{\alpha I}$ are Yukawa couplings to active lepton flavors $\alpha=e,\mu,\tau$; $L_{\alpha}=(\nu_{\alpha}, l_\alpha)^{T}$ is the lepton left doublet; and $H$ is the Higgs doublet. 

Due to the mixing with heavy sterile neutrinos, the active neutrino flavor eigenstates become
\begin{equation}
    \nu_{\alpha}= U^{\text{PMNS}}_{\alpha j} \nu_{j} + \theta_{\alpha I}N_{I}^{C},
\end{equation}
where $\nu_{j}$ and $N_{I}$ are mass eigenstates, and $U^{\text{PMNS}}$ is  the Pontecorvo–Maki–Nakagawa–\\Sakata matrix~\cite{Pontecorvo:1957qd}. The small parameters $\theta_{\alpha I}$ control the interaction strength of the HNLs with the $W$, $Z$, and $H$ bosons. Their explicit form is given by
\begin{equation}
    \theta_{\alpha I} = \frac{v}{\sqrt{2}}\sum_I F_{\alpha J} (M^{-1})_{IJ}
\end{equation}
with $v=\unit[246]{GeV}$ being the Higgs vacuum expectation value.

It is convenient to define the mixing angles, characterizing HNL interactions:
\begin{equation}
   U_{\alpha I}^2=|\theta_{\alpha I}|^2,\quad\quad U^2_{\alpha}=\sum_{I}U_{\alpha I}^2, \quad\quad  U^2_{I}=\sum_{\alpha}U_{\alpha I}^2,\quad\quad U^2=\sum_{\alpha I}U_{\alpha I}^2.
\end{equation}

At high-energy colliders, the detection of an HNL occurs by reconstructing a fully visible decay
$N\to l W \to l q q$, without a neutrino in the final state. For all relevant masses and mixing angles (see Eq.~\eqref{eq:NHNL_estimate}), the decay length cannot exceed $\mu$m, and only prompt decays can be searched for. Expressions for the decay width of right-handed neutrinos are presented in Appendix~\ref{Sub:HNL decay width}.

The total number of events can be estimated by
\begin{equation}
\label{eq:nev}
    N_\text{ev} = 2 \mathcal L \times \sum_\alpha U^2_\alpha \sigma_{N_\alpha} \times \text{Br}(N\to W(qq) l) \times \epsilon_{\text{eff}} .
\end{equation}
Here, the factor $2$ takes into account the Majorana nature of the HNLs, and $\sigma_{N_\alpha}$ represents the production cross-section of a Dirac HNL-particle (i.e., excluding charge-conjugated channels) with unit mixing angle $U^2_\beta = \delta_{\alpha\beta}$. The detection efficiency $\epsilon_{\text{eff}}$ is the probability of observing an HNL decay when it occurs. The branching ratio of the HNL decay is
\begin{equation}
\text{Br}(N\to l W \to l q q) = \text{Br}(N \to l W) \cdot \underbrace{\text{Br}(W\to qq)}_{=0.676}.
\end{equation}

In the TeV mass range, the branching ratio of HNL decay into a $W$ boson approaches $\text{Br}(N \to l W) \approx \frac{1}{2}$.

\subsection{$\mu$-mixing}
\label{sec:mumixing}

\begin{figure}[h!]
\centering
    \includegraphics[width = 0.3\textwidth]{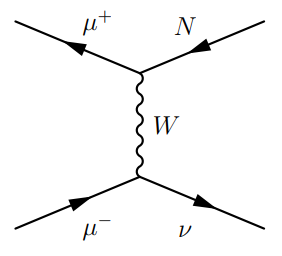}
\caption{ Feynman diagram for $\mu^{-}\mu^{+}\to \nu N$ with $W$-boson exchange in the $t$-channel.}
\label{fig:Wt-diagram}
\end{figure}

The dominant production channel is mediated through the mixing $U^2_\mu$ with the muon neutrino, the diagram is shown in Fig.~\ref{fig:Wt-diagram}.
The cross-section of production of an HNL with mass $m_N$ is given by 
\begin{equation}
    \sigma_{N_\mu} \approx \frac{g_W^4}{16\pi m_W^2} \left(1 - \frac{m_N^2}{s}\right) = \underbrace{\sigma_{\text{weak}}}_{\sim \unit[200]{pb}} \left(1 - \frac{m_N^2}{s}\right) .
\end{equation}

The production in the collinear process comes at the cost that the produced particles move at small angles to the axis, which may suppress the detection efficiency. Suppose we assume that the detector system covers angles above $\theta_\text{det}$. In that case, the detection efficiency can be estimated by the fraction of HNL whose decay products can deviate by angles $\theta > \theta_\text{det}$ from the initial beam axis. We choose the reference value $\theta_\text{det} =10^\circ$ (corresponding to the pseudo-rapidity cut $\eta = 2.44$). There are two limiting cases for the calculation of efficiency:
\begin{enumerate}
    \item If HNLs are not too light, they are produced with small boosts, and their decay products are not focused on the beam line. The typical opening angle between the decay products of a boosted particle may be estimated as $\theta_\text{dec.} \sim 1/\gamma_N$, where $\gamma_N$ is the gamma factor of the HNL. The detection efficiency, that is, the probability of emitting decay products in the detector coverage, is:
    \begin{equation}
        \label{eq:deteff1}
        \epsilon_\text{eff} \sim 1 - \left(\frac{\theta_\text{det}}{\theta_\text{dec.}}\right)^2 \sim 1 - \theta_\text{det}^2 \frac{4 s m_N^2}{(s + m_N^2)^2}.
    \end{equation}
    \item If HNLs are sufficiently light, such that $1/\gamma_N < \theta_\text{det}$, the decay products do not deviate significantly from the initial direction of the HNL motion. Therefore, it becomes necessary to account for the nonzero angle of the HNLs itself, easily computed from the angular dependence of the cross-section $\frac{d\sigma}{dt} \propto (m_W^2 - t)^{-2}$ with the requirement of HNL deflection by an angle larger than $\theta_\text{det}$. 
    \begin{equation}
        \label{eq:deteff2}
        \epsilon_\text{eff} \sim \frac{\sigma_{N \nu}[\theta > \theta_\text{det}]}{\sigma_{N\nu}} =  \left[1 + \frac{s}{m_W^2} \left(1 - \frac{m_N^2}{s}\right) \sin^2\left(\frac{\theta_\text{det}}{2}\right)\right]^{-1}.
    \end{equation}
    It should be noted that this effect becomes relevant in the regime when $m_N^2\ll s$. Therefore, the detection efficiency approaches a constant value independent of HNL mass. This approximation might be inaccurate for very light HNLs, where the separation of the decay products and reconstruction of the kinematics become challenging. Hence, we restrict ourselves to $m_N>\unit[200]{GeV}$.
\end{enumerate}

In our discussion, we omit direct cuts on the $p_T$ of the decay products because in both scenarios it is limited by $p_T \gtrsim \frac{\sqrt{s}}{2}\theta_\text{det} > \unit[100]{GeV}$. Therefore, we approximate the detection efficiency with the two simple formulas~\eqref{eq:deteff1}, \eqref{eq:deteff2}, keeping the maximal of the two.
\begin{equation}
    \label{eq:deteff}
    \epsilon_{\text{eff}}(m_N| s, \theta_\text{det}) = \text{max} \left[1 - \theta^2_\text{det} \frac{4 s m_N^2}{(s+m_N^2)^2}, \left(1 + \theta^2_\text{det}\frac{s}{4m_W^2}\right)^{-1}\right].
\end{equation}

\begin{figure}[h!]
    \centering
    \includegraphics[width = 0.5\textwidth]{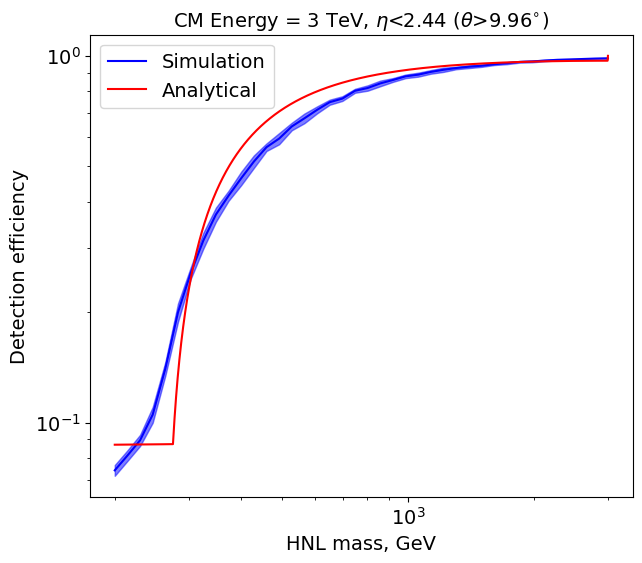}~
    \includegraphics[width = 0.5\textwidth]{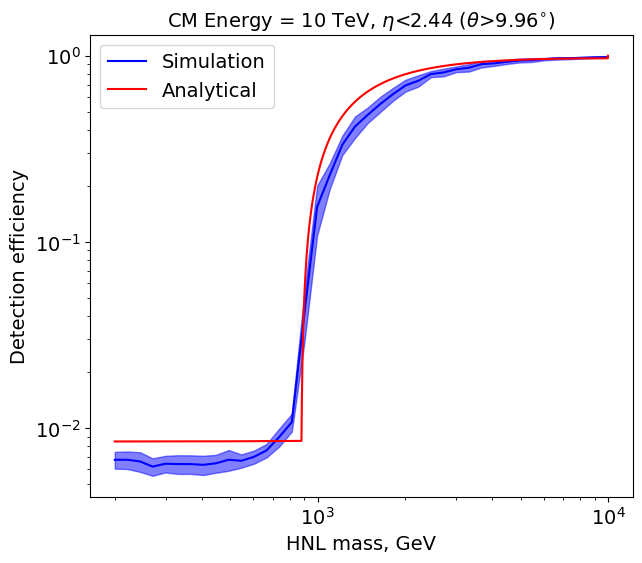}

    \caption{Comparison of the detection efficiency, computed with a simple formula~\eqref{eq:deteff}, and the results of a toy Monte-Carlo generator. The generator simulates HNL production and decay into two massless particles and estimates the decay fraction with both daughter particles moving into the detector. }
    \label{fig:eff_comparison}
\end{figure}

To verify our estimates, we made a toy Monte-Carlo generator of the decay of an HNL into two massless particles. The HNL is produced at various angles $\theta$ with the weights proportional to $\frac{d\sigma}{d\cos\theta}$. The decay products are created in HNL's reference frame and boosted to the lab frame. The fraction of decays where both decay products have $\theta > \theta_\text{det}$ is shown in Fig.~\ref{fig:eff_comparison}, together with the estimate~\eqref{eq:deteff}.

To estimate the sensitivity of an experiment at the muon collider, it is necessary to know the number of expected SM background events $N_{\text{bkg}}$. Then, the exclusion region (at $1\sigma$ confidence level) corresponds to the expected number of HNL events
\begin{equation}
    N_\text{ev} = \sqrt{N_\text{bkg}}.
\end{equation}

The number of background events lacks a simple analytical estimate. Therefore, we extract it from the simulations performed in the previous studies. For this purpose, we used the results of the previous simulation-based studies~\cite{Kwok:2023dck, Li:2023tbx, Mekala:2023diu}. By fitting the background spectrum from \cite{Kwok:2023dck} and choosing 100 GeV bins, equal to the size of a peak for the reconstructed mass, we found that for a 3 TeV muon collider $N_\text{bkg}$ can be approximated as 
\begin{equation}
    \label{eq:bkg3TeV}
    N_{\text{bkg}}=4.0\cdot10^{4}\left(\exp\bigg[-1.6\cdot \frac{m_{N}}{ \unit[1]{TeV}}\bigg] + 0.03\right).
\end{equation}
 
To validate our estimates, we compared (\ref{eq:bkg3TeV}) with the results from \cite{Li:2023tbx} and \cite{Kwok:2023dck}. Our calculations indicate roughly $\sim 10^{4}$ events in the 100 GeV bin.

For the 10 TeV collider, the background is dominated by $\mu^{+}\mu^{-}\rightarrow qq lll \nu$ for $m_{N}\;\lesssim\;5$ TeV and $\mu^{+}\mu^{-}\rightarrow qql\nu$ for $m_{N}\;\gtrsim\;5$ TeV. Additionally, there is a contribution from $\mu^{+}\mu^{-}\rightarrow qq ll$, uniform in the invariant mass. To determine the background in this scenario, we used the ratio between cross sections of those three processes (0.40 pb, 0.21 pb, and 0.46 pb correspondingly) to the total cross section from \cite{Li:2023tbx} assuming a linear $N_{\text{bkg}}(m_{N})$ dependence. Our derived equation is 
\begin{equation}
    \label{eq:bkg10TeV}
    N_{\text{bkg}}=1.5\cdot10^{5}\:\left(1 -0.07\cdot\frac{m_{N}}{\unit[1]{TeV}}\right).
\end{equation}
In addition, we performed a check with \cite{Kwok:2023dck} and \cite{Mekala:2023diu}. Here, we noticed a disagreement between the latter and previously mentioned papers, where the background is ten times larger. Unfortunately, we are unable to explain this discrepancy.

\begin{figure}[h!]
    \centering
    \includegraphics[width = 0.5\textwidth]{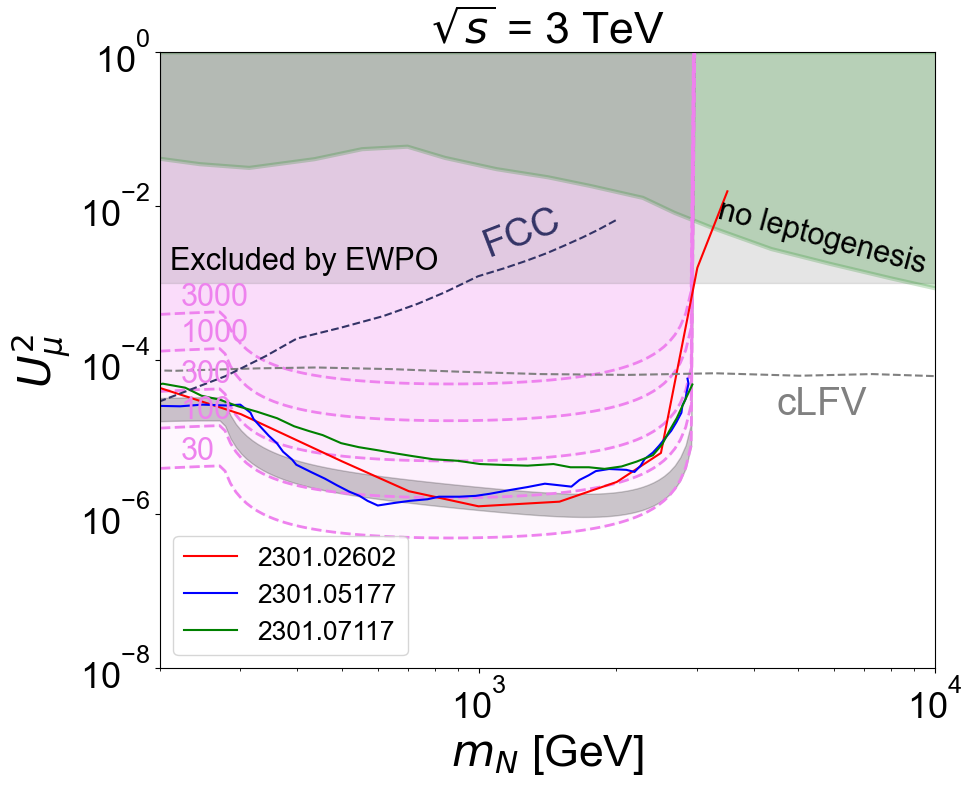}~
    \includegraphics[width = 0.5\textwidth]{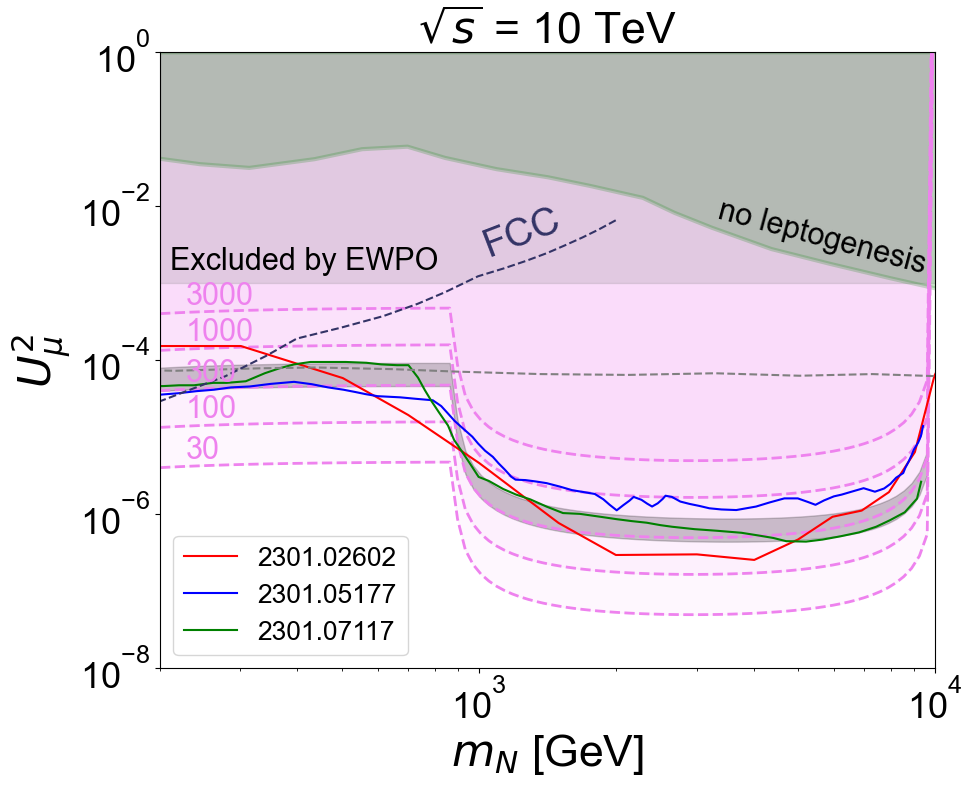}
    \caption{The $N_\text{ev}$ map in the $(m_N, U^2)$ parameter space. The black band represents the sensitivity reach using the background fits~\eqref{eq:bkg3TeV}, \eqref{eq:bkg10TeV}, allowing them to vary by a factor of two. The gray area at the level $\sim 10^{-3}$ is excluded by constraints from the electroweak precision observables~\cite{delAguila:2008pw, Chrzaszcz:2019inj}. The FCC sensitivity is taken from~\cite{Antusch:2016ejd}. The current charge lepton flavor violation constraints (cLFV)~\cite{Alonso:2012ji, Abada:2015oba,deGouvea:2015euy, Fernandez-Martinez:2016lgt, Abada:2018nio, Urquia-Calderon:2022ufc} may in principle exclude mixing angles up to $10^{-4}$. However, they rely on the simultaneous mixing of a HNL with electron and muon neutrinos. While not directly probing $U^2_\mu$, these constraints are nevertheless relevant for realistic HNL scenarios.
    }
    \label{fig:sensitivity}
\end{figure}

The sensitivity to HNLs obtained by the presented simple estimates is shown in Fig.~\ref{fig:sensitivity}. Given the somewhat arbitrary estimates for the background, we plot the sensitivity as a band, allowing for variation in the background by a factor of two. We note that the consistency of our estimates with the previous results is at the same level as the (dis)agreement between them.

\subsection{Feasibility of vector boson fusion for $e$, $\tau$-mixings}

\begin{figure}[h!]
 \centering
    \includegraphics[width = \textwidth]{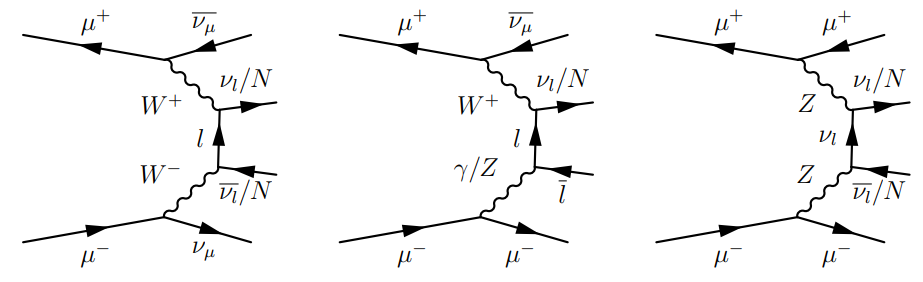}
\caption{Feynman diagrams of HNL production through vector boson fusion.}
\label{fig:VBF}
\end{figure}

A natural process that may experience collinear enhancement of the cross-section and produce HNL through mixing with $\nu_e$/$\nu_\tau$ is vector boson fusion (VBF), see Fig.~\ref{fig:VBF}. This type of interaction has been identified as a promising process
for the muon collider to probe new physics~\cite{Costantini:2020stv,  Ruiz:2021tdt, Li:2023lkl}. In this section, we demonstrate that this process does not provide an advantage compared to the $s$-channel annihilation process into $N\nu$. Moreover, we highlight the caveats of the effective $V$ approximation when considering HNL production.

\vspace{5mm}

The relevant reactions for HNL production are $W^-W^+\to \nu N$, $ZW^\pm\to l^\pm N$, $\gamma W^\pm\to l^\pm N$, and $ZZ \to \nu N$, shown in Fig~\ref{fig:VBF}. Their cross-sections can be estimated using the effective $V$ approximation. The two necessary ingredients of the computations are the probability distribution $f_V(\xi, Q^2)$ to emit a vector boson with energy fraction $\xi \equiv \frac{E_V}{E_\mu}$ and virtuality $q^2 = - Q^2 < 0$, and the fusion cross section $\sigma(\hat s, Q_1^2, Q_2^2)$, with the center of mass energy $\hat s = \xi_1 \xi_2 s$. A detailed treatment of the emission of $V$ may be found elsewhere, see e.g.~\cite{Costantini:2020stv, Borel:2012by, Kuss:1995yv, Dawson:1984gx, Kane:1984bb}.

For our purposes, we capture the general behavior of the emission probability by approximating
\begin{equation}
    f_V(\xi, Q^2) \sim \frac{g_V^2}{16\pi^2} \frac{1}{\xi} \frac{1}{Q^2 + m_V^2},
\end{equation}
which results in
\begin{equation}
\label{eq:sigma_vbf}
    \sigma_{\text{VBF}} \sim \frac{g_V^4}{256\pi^4} \int \frac{d\xi_1}{\xi_1} \frac{d\xi_2}{\xi_2} \frac{dQ_1^2}{Q_1^2 + m_V^2} \frac{dQ_2^2}{Q_2^2 + m_V^2} \times \theta(\hat s - m_N^2) \sigma(\hat s, Q_1^2, Q_2^2) 
\end{equation}
with $\theta$ being the Heaviside step-function.

The next step of the effective $V$ approximation is to replace the off-shell cross section by its value for on-shell vector bosons $\sigma(\hat s, -m_V^2, -m_V^2)$. However, this step becomes invalid for the specific process of HNL production. For instance, consider a process $W^-W^+ \to \nu N$. The total cross-section for real $W$ bosons is divergent because of the $t$-channel singularity. Namely, the range of variable $t$ in the limit $m_W^2 \ll m_N^2, s$ is
\begin{equation}
    \label{eq:t_range}
    m_W^2- s\left( 1 - \frac{m_N^2}{s}\right)  \leq t \leq m_W^2\left[1  - \frac{1}{4}\left( 1 - \frac{m_N^2}{s}\right)\right].
\end{equation}
Therefore, this variable crosses the zero value, i.e., contains an on-shell lepton in the $t$-channel propagator, for HNL masses
\begin{equation}
     m^2_N < s - m_W^2,
\end{equation}
The singularity appears for almost all HNL masses except for a narrow region near the kinematic threshold $s\approx m_N^2$.

This type of singularity has been first noticed in the context of hadron collisions~\cite{peierls1961possible, Li:1997zb, Li:1997bi}, and has received attention for applications in cosmology~\cite{Grzadkowski:2021kgi, Iglicki:2022jjf} and the context of the processes at a muon collider~\cite{ginzburg1995effect, ginzburg1996initial, melnikov1997processes, dams2003singular}. It had been suggested to regularize the singularity by accounting for the decay width of the incoming particles or the transversal size of the beam in the case of the muon collider. Both suggestions rely on the fact that the initial state particles cannot be treated as well-defined plane waves. This violation of the energy-momentum conservation prevents the particle in the propagator from being exactly on-shell.

It is essential to recognize that in the context of HNL production, the singularity cannot be resolved by implying kinematic cuts on the final particles. As discussed in Sec.~\ref{sec:mumixing}, sufficiently heavy HNLs may be produced with small boosts, and their decay products can be detected with $\approx 1$ efficiency. The critical parameter is the total number of HNLs created in the process. 

Therefore, the cross-section $\sigma(\hat s, Q_1^2, Q_2^2)$ should be evaluated strictly for off-shell vector bosons with negative invariant mass $m_W^2\to -Q^2$. In this case, there is no zero crossing in the $t$ range~\eqref{eq:t_range}. Using the unitarity constraints that forbid $\sigma$ from growing with energy, we can estimate the upper bounds on the cross-section by:
\begin{equation}
\label{eq:ranges}
    \sigma(s, Q_1^2, Q_2^2) \lesssim \frac{g^4}{16\pi}
    \begin{cases}
        \frac{1}{m_W \Gamma_W}, & Q^2\lesssim m_W \Gamma_W , \\
        \frac{1}{Q^2}, & m_W \Gamma_W \lesssim Q^2\lesssim m_W^2 , \\
        \frac{1}{m_W^2}, & m_W^2 \lesssim Q^2 ,
    \end{cases}
\end{equation}
where $Q^2$ represents any of the quantities $Q_1^2$, $Q_2^2$. The estimate for the case $Q^2\lesssim m_W \Gamma_W$ accounts for the instability of the incoming particles, essentially employing the same argument that was used for the $t$-channel singularity~\cite{ginzburg1996initial}.

The integration over $Q^2_{1,2}$ in Eq.~\eqref{eq:sigma_vbf} can be split into three ranges, listed in Eq.~\eqref{eq:ranges}. This results in the following upper bound
\begin{multline}
    \hat \sigma(\hat s) \equiv \int \frac{dQ^2_1}{Q_1^2 + m_W^2}\frac{dQ^2_2}{Q_2^2 + m_W^2} \sigma(\hat s, Q_1^2, Q_2^2) \lesssim  \\
    \lesssim \frac{g^4}{16\pi}\left( \pi \frac{(m_W \Gamma_W/m_W^2)^2}{m_W \Gamma_W} + \pi\frac{1 - \Gamma_W/m_W}{m_W^2}  + \frac{1}{m_W^2}\left(\ln \frac{\hat s}{m_W^2}\right)^2 \right),
\end{multline}
The first two terms are negligible compared to the third term, which is the usual $\sigma_\text{weak}$ cross-section enhanced by the collinear emission of the vector bosons. Therefore, the final VBF cross-section for HNL production~\eqref{eq:sigma_vbf} is bounded by
\begin{equation}
    \sigma_{\text{VBF}} \lesssim \sigma_\text{weak} \times \frac{g^4}{256\pi^4} \times \left(\ln \frac{s}{m_W^2}\right)^2 \times \ln\frac{s}{m_N^2} \approx 10^{-3} \times \sigma_\text{weak}.
\end{equation}

For comparison, the cross-section of $s$-channel production is
\begin{equation}
    \sigma_s \sim \frac{m_W^2}{s} \sigma_\text{weak} = (10^{-4} \,\mbox{---}\,10^{-3} ) \times \sigma_\text{weak}
\end{equation}
for the center-of-mass energies $\sqrt{s} = 3$ -- $\unit[10]{TeV}$.

Therefore, we conclude that the account for the VBF processes does not provide a significant improvement in sensitivity compared to the $s$-channel. Moreover, in the most optimistic scenario, the sensitivity contour can reach $U^2_\alpha \sim 10^{-3}$, while the current constraints on $e$ and $\tau$ mixing are $U^2_e\lesssim 10^{-3}$ and $U_\tau \lesssim 10^{-2}$~\cite{Chrzaszcz:2019inj}.

At the same time, vector boson fusion may become the dominant process for the production of $\nu_e$- and $\nu_\tau$-mixing HNLs at muon colliders of much higher energy above $\unit[10]{TeV}$. An accurate analysis of the effective boson approximation would be required in this case.

\section{Probing lepton number violation and leptogenesis}
\label{Sec:3}

The muon collider is a perfect instrument to measure the Majorana vs. Dirac nature of the HNL. Similar studies at various collider and beam-extracted experiments have been performed in~\cite{Tastet:2019nqj, Antusch:2016ejd, Drewes:2022rsk, Antusch:2023jsa, Deppisch:2013jxa}. The specific advantage of the muon collider stems from the fact that the HNL momentum tends to point in the same forward direction $\theta <\pi/2$ as the muon it originated from.

Each reconstructed signal event can be classified by the initial lepton number, which is labeled according to the corresponding hemisphere containing the total momentum of the decay products. This lepton charge is then compared with the charge of the produced muon in the $N\to \mu qq$ decay to eventually label the process as lepton number violating (LNV) or conserving (LNC).

\subsection{Sensitivity to lepton number violation}

To quantify the measure of lepton number violation, we define the parameter $A$, being twice the probability that HNL, produced in the mixing with a lepton, decays into the opposite-charge lepton:
\begin{equation*}
    A = 2 \times \text{Br}(N_l \to \bar l\dots).
\end{equation*}
With this definition, a Dirac HNL corresponds to $A = 0$, while for a purely Majorana HNL $A=1$. 

Given the total number of detectable HNL decays $N_\text{ev}$, the counts of observed LNV-like and LNC-like events contain three contributions: from correctly and incorrectly classified signal events and background events:
\begin{align*}
    &N_\text{LNV} = \frac{A}{2}\cdot(1 - f) N_\text{ev} + \left(1-\frac{A}{2}\right)\cdot f N_\text{ev} +  N_\text{LNV, bkg}, \\
    &N_\text{LNC} = \left(1 - \frac{A}{2}\right) \cdot (1-f)N_\text{ev} + \frac{A}{2}\cdot f N_\text{ev} + N_\text{LNC, bkg} ,
\end{align*}
Here, $f$ is the probability of incorrect classification of the reconstructed HNL decay, $N_{\text{LNV/C, bkg}}$ are contributions of the SM background to each event class.

Incorrect signal classification may happen due to backward production of an HNL at large angles $\theta >\pi/2$, and because of poor reconstruction of the total momentum. The probability $f_{B}$ of backward production ($\theta>\pi/2$) can be approximated numerically as 
\begin{equation}
    f_{B} \approx 0.07 \frac{m_W^2}{s} \times \frac{(1+m_N^2/s)^2}{1 - m_N^2/s} \approx 5\cdot 10^{-5}\,\frac{\unit[3]{TeV}^2}{s}, \quad s-m_N^2 \gg m_W^2,
\end{equation}
where the prefactor is computed by accounting for both $W$ and $ Z$-boson-mediated processes. The total momentum of the HNL is $p_N \sim \frac{1}{2}\sqrt{s} (1 - m_N^2/s)$. Assuming that the uncertainty of momentum reconstruction is not too large, i.e. $\lesssim m_W$, both effects are negligible in the large range of HNL masses up to the production threshold $s-m_N^2 \lesssim m_W^2$. For the sake of simplicity, we ignore this specific case and set $f=0$. 

We assume that the contribution of the SM background is symmetric for LNV/LNV-like events: $N_{\text{LNV, bkg}} = N_{\text{LNC, bkg}} = N_{\text{bkg}}/2$. This assumption needs to be verified by background simulations, given that the SM processes with the initial muons may be more likely to produce the same sign lepton in the forward direction. In this case, the background for the LNV-like events becomes lower, improving the overall sensitivity. Therefore, we consider our assumption conservative.  

The sensitivity to $A$ is estimated using the $\chi^2$ test. For a model that is parametrized $(N_\text{ev}, A)$, we may test an alternative model $(\tilde N_\text{ev}, \tilde{A})$ by defining the quantity
\begin{equation}
    \chi^2(\tilde N_\text{ev}, \tilde{A} | N_\text{ev}, A, N_\text{bkg}) = \frac{(\tilde N_{\text{LNV}} - N_{\text{LNV}})^2}{\tilde N_\text{LNV}} + \frac{(\tilde N_{\text{LNC}} - N_{\text{LNC}})^2}{\tilde N_\text{LNC}} .
\end{equation}

\begin{figure}[h!]
    \centering
    \includegraphics[width = \textwidth]{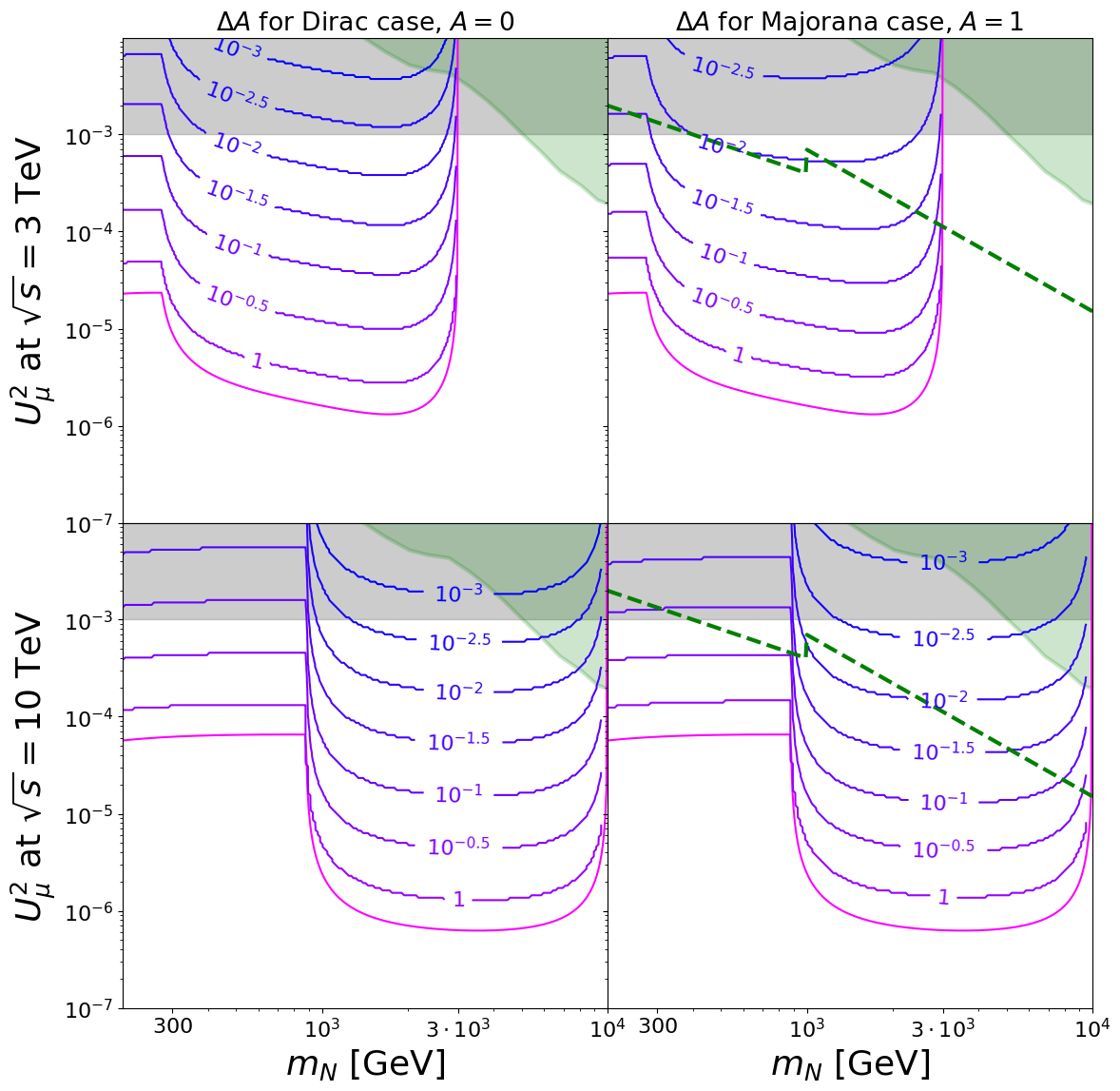}

    \caption{Constraints on lepton number violation $\Delta A$ deviations for the pure Dirac $A=0$ (left) and Majorana $A = 1$ (right) cases for the $\sqrt{s} = \unit[3]{TeV}$ (top) and $\sqrt{s} = \unit[10]{TeV}$ (bottom) muon collider. The gray area is the excluded region. The green shaded area represents the parameter space where successful leptogenesis with neither thermal nor vanishing initial conditions is possible~\cite{Drewes:2021nqr}. The green dashed line represents the scale above which the 3 HNLs in the Majorana-like case would produce large loop corrections to active neutrino masses~\cite{Drewes:2019byd}.}
    \label{fig:A_sensitivity}
\end{figure}

We define the precision of the $A$ determination as the interval $\Delta A$ of variation of $\tilde A$, in which $\chi^2(\tilde A)$ does not exceed the $1\sigma$ quantile of the chi-squared distribution with 2 degrees of freedom $\chi^2 < 2.30$, after marginalization over $\tilde N_\text{ev}$. Using the estimates for the number of signal and background events, we show this sensitivity in Fig.~\ref{fig:A_sensitivity}.

The numerical results for the precision $\Delta A$ of the determination of $A$ may be approximated by a simple relation
\begin{equation}
    \Delta A \approx \frac{\sqrt{N_{\text{ev}} \cdot A + N_\text{bkg}}}{N_{\text{ev}}}.
\end{equation}
If the background fluctuations dominate uncertainty, the sensitivity scales as $\Delta A \sim 1/U^2$. This relation always holds for the Dirac case $A = 0$. Conversely, once $N_\text{ev} A > N_{\text{bkg}}$, the sensitivity starts to grow with the mixing angle at a smaller pace $\propto 1/\sqrt{U^2}$.

\subsection{Implications for leptogenesis}

In the scenario with two degenerate HNLs $N_1$, $N_2$, the coupling matrix $\theta_{\alpha I}$ and mass matrix has the form
\begin{equation}
    \theta_{\alpha I} = 
    \frac{1}{\sqrt{2}}
    \begin{pmatrix} 
        U_e & i U_e
        \\
        U_\mu & i U_\mu
        \\
        U_\tau & i U_\tau
    \end{pmatrix},
    \qquad \qquad 
    M_{IJ} = 
    \begin{pmatrix}
        m_N- \frac{\Delta M}{2} & 0 \\
        0 & m_N + \frac{\Delta M}{2} 
    \end{pmatrix}.
\end{equation}

The two HNLs form a quasi-Dirac pair, $\Psi = \frac{1}{\sqrt{2}}(|N_1\rangle + i |N_2 \rangle)$, $\bar \Psi = \frac{1}{\sqrt{2}}(|N_1\rangle - i |N_2 \rangle)$. This pair violates the lepton charge only due to the relative oscillations between the HNLs arising from a slight difference in their mass $\Delta M$.

The expression for the parameter $A$ has the form:
\begin{equation}
    A = 2 \frac{\int_0^\infty |g_-(t)|^2 \, dt }{\int_0^\infty (|g_+(t)|^2 + |g_-(t)|^2) dt},
\end{equation}
where $g_{\pm}$ are the matrix elements for LNC/LNV transitions, defined as
\begin{equation}
        g_+(t) = \Psi^\dagger \mathcal U(t)\Psi, \qquad g_-(t) = \bar \Psi^\dagger \mathcal U(t) \Psi,
\end{equation}
where $U(t)$ is the evolution matrix. In the 2 HNL model, it has the form
\begin{equation}
    \mathcal U(t) = \exp\left(-i t M_{IJ} - \frac{\Gamma}{2} t\right),
\end{equation}
and the lepton number violation parameter is given by the ratio~\cite{Anamiati:2016uxp, Deppisch:2015qwa}:
\begin{equation}
    \label{eq:2HNL_A}
    A =  \frac{\Delta M^2}{\Gamma^2 + \Delta M^2}.
\end{equation}

In the 3 HNL scenario, the coupling and mass matrix can be rewritten as
\begin{equation}
    \label{eq:3HNLmatrices}
    \theta_{\alpha I} = 
    \frac{1}{\sqrt{2}}
    \begin{pmatrix} 
        U_e & i U_e & 0
        \\
        U_\mu & i U_\mu & 0
        \\
        U_\tau & i U_\tau & 0
    \end{pmatrix},
    \qquad \qquad 
    M_{IJ} = m_N\cdot \mathbbm 1_{3\times 3} +\Gamma \begin{pmatrix}
        0 & 0 & \xi_1\\
        0 & \mu_2 & \xi_2 \\
        \xi_1 & \xi_2 & \mu_3
    \end{pmatrix},
\end{equation}
assuming that the couplings to the active neutrino sector explicitly respect the lepton symmetry. All deviations in the mass matrix are defined as normalized to the decay width. In this case, the evolution matrix has the form
\begin{equation}
    \mathcal U(t) = e^{-im_Nt} \exp\left[-\Gamma t 
    \begin{pmatrix} 
        \frac12 & 0 & i\xi_1 \\
        0 & \frac12+ i \mu_2 &  i\xi_2 \\
        i\xi_1 & i\xi_2 & i\mu_3
    \end{pmatrix}
    \right].
\end{equation}

The larger number of independent parameters adds two difficulties in relating the LNV parameter to the underlying model. First, a simple analytic expression of the form~\eqref{eq:2HNL_A} is missing, and in general, $A$ should be evaluated numerically. Second, the measurement of $A$ adds only one constraint to the set of splitting parameters $\{\mu_i$, $\xi_i\}$, insufficient for reconstructing all the parameters.

To quantitatively explain the way LNV occurs, we need to separate the two main effects. If the oscillations between the interacting states $|N_1\rangle$ and $|N_2\rangle$ are rapid enough $\mu_2 \gtrsim 1$, we return to the case of 2 HNLs that exhibit Majorana-like behavior $A\sim 1$. If, in contrast, these oscillations are suppressed $\mu_2 \ll 1$, the third HNL starts to play a role. If $|\xi|\gg |\mu_3|$, the massive states of sterile neutrinos become entirely misaligned with the initial interacting states, and the mass splittings become of order $\sim \Gamma\xi$. Finally, if $|\xi|\ll |\mu_3|$, a situation similar to the seesaw mechanism occurs: the interacting states $\Psi$ acquire a small admixture of the third HNL with an amplitude of order $\frac{\xi}{\mu_3}$. Then the production and decay processes with a Dirac-like HNL with mixing angles $U^2$ are accompanied by the processes with a Majorana HNL $|N_3\rangle$ with mixing angles suppressed to $U^2 \cdot \frac{\xi}{\mu_3}$.

\begin{figure}[h!]
    \centering
    \includegraphics[width = 1\textwidth]{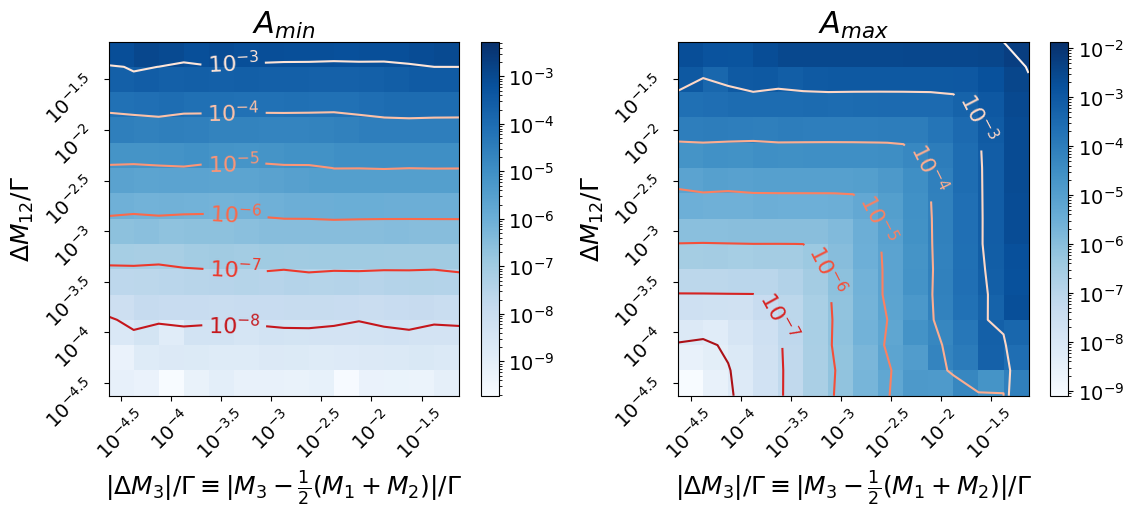}
    
    \caption{The minimal and maximal values of $A$ for given pair of mass splittings $\Delta M_{12}$ and $\Delta M_3$, see text for conventions. }
    \label{fig:DmAsl}
\end{figure}

If the parameters that define the scale of the probability of LNV to happen are sufficiently small $|\mu_2|$, $|\xi| \ll 1$ (but arbitrary $\mu_3$), the leading term expansion may be written explicitly:
\begin{equation}
    \label{eq:A_small}
    A = \mu_2^2 + \frac{2(\xi_1^2+\xi_2^2)}{1+4\mu_3^2}.
\end{equation}
In general, this expression cannot be directly linked to the mass splittings. To fix the convention, we define the ``third'' massive state with mass $M_3$ as the eigenstate of the mass matrix that is closest to the initial third state in the basis of Eq.~\eqref{eq:3HNLmatrices}, and the first two as $M_2 > M_1$. Then, there are two mass splittings $\Delta M_{12} \geq 0$ and $\Delta M_3 = M_3 - \frac{M_1+M_2}{2}$. To show the variation in $A$ for various combinations of masses, we performed a scan in the parameters $\{\mu_i$, $\xi_i\}$. The minimal and maximal values of $A$ that can be achieved for given $\Delta M_{12}$, $|\Delta M_3|$ are shown in Fig.~\ref{fig:DmAsl}.

While $A$ is not fixed for given mass splittings, one can estimate the order of magnitude of the mass splittings $\Delta M \sim \Delta M_{12}, |\Delta M_3|$.
\begin{equation}
\label{eq:dmm}
    A \sim \frac{\Delta M^2}{\Gamma^2}: \qquad
    \frac{\Delta M}{m_N} \sim \frac{\Gamma(m_N) \sqrt{A}}{m_N} \approx \sqrt{A}\, U^2 \left(\frac{m_N}{\unit[1]{TeV}}\right)^2, \quad m_N \gtrsim \unit[200]{GeV}  .
\end{equation}

This relation has been confirmed by a numerical scan, which shows that the approximate variation range of the parameter $A$ is $[0.5 \Delta M^2_{12}, \text{max}(\Delta M^2_{12}, \Delta M^2_3)]/\Gamma^2$ for the mass splittings that differ less by an order of magnitude.

\begin{figure}[h!]
    \centering
    \includegraphics[width = \textwidth]{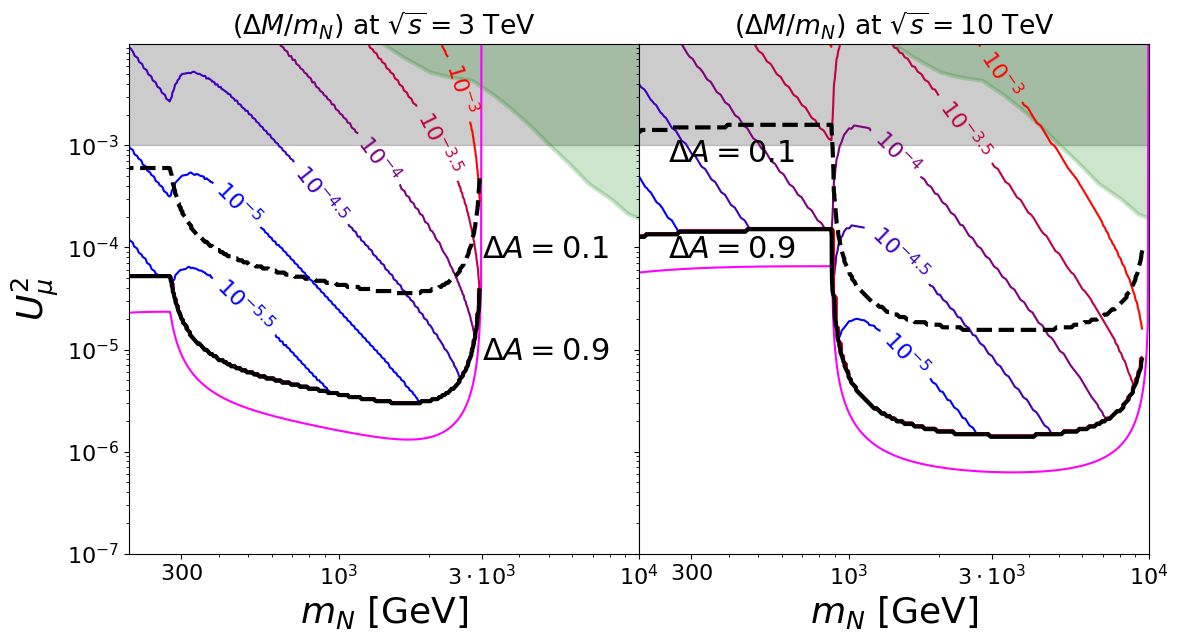}  
    \caption{The lower bounds on the scale of mass splitting $\Delta M/m_N$ (defined as Eq.~\eqref{eq:dmm}) that may be probed by the LNV measurements at a muon collider, assuming an approximately Dirac-like scenario $A\ll 1$. The black dashed line represents the sensitivity $\Delta A = 0.1$, for which measurements of the mass splitting via~\eqref{eq:dmm} is possible. Below it down to $\Delta A= 0.9$ (black solid line), the precision of the $A$ measurement is insufficient, and only a preference towards the Majorana/Dirac nature can be established. In this region, $(\Delta M/m_N)$ represents the transition point between the two cases. }
    \label{fig:dMM_sensitivity}  
\end{figure}

Given these relations, for a specific measurement of $A$, one can reduce the parameter space of the models with 3 HNL, which may explain the baryon asymmetry of the Universe. A scan of the values of $A$ with a mapping onto the sensitivity map of the muon collider experiments may reveal some potential benefits of this measurement. The lower bounds of $\Delta M / m_N$ defined by Eq.~\eqref{eq:dmm} that may be probed by a muon collider (replacing $A$ by $\Delta A$) are shown in Fig.~\ref{fig:dMM_sensitivity}.

 \section{Conclusions}
 \label{Sec:conclusion}
In this paper, we examined the potential of future muon colliders to measure lepton number violation in searches for heavy neutral leptons. This probe may address the role played by the newfound particles in explaining the matter-antimatter imbalance. The sensitivity of the muon collider down to $U_\mu^2 \sim 10^{-6}$ provides a unique probe of the TeV-scale HNL, superior to the competing FCC, CEPC, CLIC, and ILC projects. It enters the parameter space where these sterile neutrinos might account for the observed value baryon asymmetry. 

We discussed the procedure for measuring lepton number violation in HNL decays and estimated the sensitivity to such searches. With the potential to observe up to $10^5$ events in the unexplored region, an experiment at the muon collider affords a sensitivity to the ratio of lepton number violating decays that can reach a percent level, as shown in Fig.~\ref{fig:A_sensitivity}. This sensitivity is crucial for deciphering the true nature of the HNL --- be it Majorana, Dirac, or situated in a non-minimal framework involving multiple nearly degenerate heavy neutral leptons.

In the sensitivity range of the muon collider, the minimal scenario of leptogenesis with two HNL fails to produce the observed matter-antimatter imbalance, and three degenerate species are needed. We relate the properties of these HNLs to the lepton number violation parameter $A$. We showed the absence of a direct link between the LNV ratio and the mass splittings, contrary to the minimal case of two HNLs. This distinction is attributed to more parameters responsible for breaking the approximate lepton symmetry. Nevertheless, the overall scale of the largest mass splitting between the HNLs is bounded by the measurements of LNV processes. The most stringent lower bounds on the $(\Delta M/m_N)$ ratio that could be attained at the muon collider are shown in Fig.~\ref{fig:dMM_sensitivity}.

The values of $\Delta M/M$ required for successful leptogenesis with three TeV-scale HNLs are missing in the literature. At the GeV scale, the relative mass splittings can vary by orders of magnitude from 1 down to $10^{-10}$~\cite{Abada:2018oly}. Our findings underscore the need for a joint analysis to establish whether the parameter space of possible models can be probed by future experimental searches. 

Lastly, we complemented the previous simulation-based estimates of the sensitivity reach of an experiment at the muon collider to sterile neutrinos. Moreover, we demonstrated the failure of the vector boson fusion processes to probe couplings to electron- and tau-neutrinos beyond the excluded bounds, as well as highlighted theoretical nuances accompanying the description of these processes.

\textbf{Note:} after the completion of the analysis, we became aware of the decoherence effects in HNL oscillations~\cite{Antusch:2023nqd}, which may drastically modify the relation of the LNV parameter to the mass splittings. These effects depend on the details of the experimental resolution and require an additional study.

\section*{Acknowledgements}
We thank Juraj Klaric for discussing the leptogenesis in the model of 3 HNLs and for proofreading the manuscript. The evaluation of the cross sections has been done with the use of FeynCalc~\cite{Shtabovenko:2016sxi, Shtabovenko:2020gxv}. This project has received funding from the European Research Council (ERC) under the European Union's Horizon 2020 research and innovation programme (GA 694896), from the NWO Physics Vrij Programme ``The Hidden Universe of Weakly Interacting Particles'', No. 680.92.18.03, which is partly financed by the Dutch Research Council NWO, and from the Den Adel Fund in the form of a scholarship.

\appendix

\section{HNL decay width}
\label{Sub:HNL decay width}

There are three decay channels for HNLs. Their partial decay widths are \cite{gluza1997heavy, del2006neutrino}:
\begin{equation}
         \Gamma(N\rightarrow W l_\alpha) 
         =\frac{g^{2}}{64\pi}U^2_\alpha \frac{m_{N}^{3}}{M_{W}^{2}}\bigg(1-\frac{M_{W}^{2}}{m_{N}^{2}}\bigg)^2\bigg(1+2\frac{M_{W}^{2}}{m_{N}^{2}}\bigg),
\end{equation}
\begin{equation}
         \Gamma(N\rightarrow Z\nu_\alpha)
         =\frac{g^{2}}{128\pi }U^2_\alpha\frac{m_{N}^{3}}{M_{W}^{2}}\bigg(1-\frac{M_{Z}^{2}}{m_{N}^{2}}\bigg)^2\bigg(1+2\frac{M_{Z}^{2}}{m_{N}^{2}}\bigg),
\end{equation}
\begin{equation}
         \Gamma(N\rightarrow H\nu_\alpha)
         =\frac{g^{2}U^2_\alpha}{128\pi}\frac{m_{N}^{3}}{M_{W}^{2}}\bigg(1-\frac{M_{H}^{2}}{m_{N}^{2}}\bigg)^{2},
\end{equation}
and approach the ratio $\Gamma(N\to Wl):\Gamma(N\to Zl):\Gamma(N\to H\nu) = 2:1:1$ in the limit $m_N \gg M_{Z,W,H}$.

We should note that the apparent strong dependence on particle mass $\Gamma \sim U^2 m_N^3/m_W^2$ is an artifact of the parametrization. In terms of the initial Yukawa couplings $F$ of the Lagrangian~\eqref{eq:HNL_Lagrangian}, the scaling is $\Gamma \sim F^2 m_N$. The perturbative unitarity constraint $F\lesssim\sqrt{4\pi}$ assures that the decay width does not exceed half the particle mass~\cite{Chanowitz:1978mv, Korner:1992an, Ilakovac:1999md}. The corresponding constraint on the mixing angle becomes relevant for HNL masses only above $\unit[10^6]{GeV}$.

\bibliographystyle{JHEP}
\bibliography{bib}

\end{document}